\documentclass{amsart}
\usepackage{hyperref}

\newtheorem{theorem}{Theorem}[section]
\newtheorem{lemma}[theorem]{Lemma}
\newtheorem{corollary}[theorem]{Corollary}
\newtheorem{remark}[theorem]{Remark}
\newtheorem{hypo}[theorem]{Hypothesis {\bf H.}\hspace*{-0.6ex}}

\newcommand{\R}{{\mathbb R}}
\newcommand{\N}{{\mathbb N}}
\newcommand{\Z}{{\mathbb Z}}
\newcommand{\C}{{\mathbb C}}
\newcommand{\M}{{\mathbb M}}

\newcommand{\nn}{\nonumber}
\newcommand{\be}{\begin{equation}}
\newcommand{\ee}{\end{equation}}
\newcommand{\bea}{\begin{eqnarray}}
\newcommand{\eea}{\end{eqnarray}}
\newcommand{\ul}{\underline}
\newcommand{\ti}{\tilde}

\newcommand{\spr}[2]{\langle #1 , #2 \rangle}

\newcommand{\re}{\mathrm{Re}}

\newcommand{\lz}{\ell^2(\Z)}
\newcommand{\tl}{\mathrm{TL}}

\newcommand{\Ker}{\mathrm{Ker}}
\newcommand{\res}{\mathrm{Res}}

\newcommand{\bpsi}{\bar{\psi}}

\newcommand{\ulz}{\ul{z}}
\newcommand{\hmu}{\hat{\mu}}

\newcommand{\dimuz}{\di_{\ul{\hat{\mu}}}}

\newcommand{\vrc}{\ul{\Xi}_{p_0}}
\newcommand{\hvrc}{\ul{\hat{\Xi}}_{p_0}}

\newcommand{\di}{\mathcal{D}}

\newcommand{\Amap}{\ul{A}_{p_0}}
\newcommand{\amap}{\ul{\alpha}_{p_0}}
\newcommand{\hAmap}{\ul{\hat{A}}_{p_0}}
\newcommand{\hamap}{\ul{\hat{\alpha}}_{p_0}}

\newcommand{\Rg}[1]{R_{2g+2}^{1/2}(#1)}

\newcommand{\sig}{\sigma}
\newcommand{\lam}{\lambda}
\newcommand{\gam}{\gamma}
\newcommand{\om}{\omega}

\numberwithin{equation}{section}

\begin{document}

\title[Soliton Solutions of the Toda Hierarchy]{Soliton Solutions of the Toda Hierarchy on
Quasi-Periodic Backgrounds Revisited}

\author[I. Egorova]{Iryna Egorova}
\address{Kharkiv National University\\ 47,Lenin ave\\ 61164 Kharkiv\\ Ukraine}
\email{\href{mailto:egorova@ilt.kharkov.ua}{egorova@ilt.kharkov.ua}}

\author[J. Michor]{Johanna Michor}
\address{Faculty of Mathematics\\
Nordbergstrasse 15\\ 1090 Wien\\ Austria\\ and International Erwin Schr\"odinger
Institute for Mathematical Physics, Boltzmanngasse 9\\ 1090 Wien\\ Austria}
\email{\href{mailto:Johanna.Michor@esi.ac.at}{Johanna.Michor@esi.ac.at}}
\urladdr{\href{http://www.mat.univie.ac.at/~jmichor/}{http://www.mat.univie.ac.at/\~{}jmichor/}}

\author[G. Teschl]{Gerald Teschl}
\address{Faculty of Mathematics\\
Nordbergstrasse 15\\ 1090 Wien\\ Austria\\ and International Erwin Schr\"odinger
Institute for Mathematical Physics, Boltzmanngasse 9\\ 1090 Wien\\ Austria}
\email{\href{mailto:Gerald.Teschl@univie.ac.at}{Gerald.Teschl@univie.ac.at}}
\urladdr{\href{http://www.mat.univie.ac.at/~gerald/}{http://www.mat.univie.ac.at/\~{}gerald/}}

\thanks{Math. Nach. (to appear)}

\thanks{Work supported by the Austrian Science Fund (FWF) under Grant
No.\ P17762 and INTAS Research Network NeCCA 03-51-6637.}


\begin{abstract}
We investigate soliton solutions of the Toda hierarchy on a
quasi-periodic finite-gap background by means of
the double commutation method and the inverse scattering
transform. In particular, we compute the phase shift caused by a soliton 
on a quasi-periodic finite-gap background. Furthermore, we consider
short range perturbations via scattering theory. We give a full description of
the effect of the double commutation method on the scattering data and
establish the inverse scattering transform in this setting.
\end{abstract}

\maketitle

\section{Introduction}

Solitons on a (quasi-)periodic background have a long tradition
and are used to model localized excitements on a phonon, lattice, or magnetic
field background (see, e.g., \cite{foml}, \cite{kumi}, \cite{mamo}, \cite{peza}, \cite{rub}, \cite{sh1},
\cite{sh2} and the references therein). Of course periodic
solutions, as well as solitons travelling on a periodic background, are well understood. 
Nevertheless there are still several open questions.

One of them is the stability of (quasi-)periodic solutions. For the constant solution
it is a classical result, that a small initial perturbation asymptotically splits in a number of
stable solitons. For a (quasi-)periodic background this cannot be
the case. In fact, associated with every soliton there is a phase shift (which will
be explicitly computed in Section~\ref{secdc}) and the phase shifts of all solitons
will not add up to zero in general. Hence there must be something which makes
up for this phase shift. Moreover, even if no solitons are present, the asymptotic limit
is not the (quasi-)periodic background!
A precise description of the asymptotic limit in terms of Abelian integrals on the
underlying Riemann surface is given in \cite{kateptr} (see \cite{katept} for a proof).
In particular, the asymptotic limit can be split into parts, one which stems from
the discrete spectrum (solitons) and one which stems from the continuous
spectrum.

The soliton part can be understood by adding/removing the solitons
using a Darboux-type transformation, that is, commutation methods for the
underlying Jacobi operators. Hence the purpose of the present paper is to
complement \cite{katept} and provide a detailed description of the
double commutation method when applied to a short range perturbation of
a quasi-periodic finite-gap solution of the Toda lattice. In particular, we are
interested in the effect of one double commutation step on the scattering data.
 
After introducing the Toda hierarchy in Section~\ref{secth} and recalling some
necessary facts on algebro-geometric quasi-periodic finite-gap solutions
in Section~\ref{secqp} we briefly review the single and double commutation methods
in Section~\ref{secdc} and compute the phase shift (in the Jacobian variety)
caused by inserting one eigenvalue for both methods. In Section~\ref{secst} we
review direct scattering theory for Jacobi operators with different (quasi-)periodic
asymptotics in the same isospectral class. As our main result we give a complete description
of the effect of the double commutation method on the scattering data. In addition,
we provide some detailed asymptotic formulas for the Jost functions $\psi_\pm(z,n)$
(which are normalized as $n\to\pm\infty$) at the {\em other} side, that is, as $n\to\mp\infty$.
Our final Section~\ref{secist} establishes the inverse scattering transform
for this setting. Our main results here are the time dependence of both the scattering data
and the kernel of the GelÕfand-Levitan-Marchenko equation.

\section{The Toda hierarchy}
\label{secth}

In this section we introduce the Toda hierarchy using the standard Lax formalism
(\cite{lax}). We first review some basic facts from \cite{bght} (see also \cite{tjac}).

We will only consider bounded solutions and hence require

\begin{hypo} \label{H t0}
Suppose $a(t)$, $b(t)$ satisfy
\[
a(t) \in \ell^{\infty}(\Z, \R), \qquad b(t) \in \ell^{\infty}(\Z, \R), \qquad
a(n,t) \neq 0, \qquad (n,t) \in \Z \times \R,
\]
and let $t \mapsto (a(t), b(t))$ be differentiable in 
$\ell^{\infty}(\Z) \oplus \ell^{\infty}(\Z)$.
\end{hypo}

\noindent
Associated with $a(t), b(t)$ is a Jacobi operator
\begin{equation}
H(t): \lz  \to  \lz, \qquad f \mapsto \tau(t) f,
\end{equation}
where
\begin{equation}
\tau(t) f(n)= a(n,t) f(n+1) + a(n-1,t) f(n-1) + b(n,t) f(n)
\end{equation}
and $\lz$ denotes the Hilbert space of square summable (complex-valued) sequences
over $\Z$. Moreover, choose constants $c_0=1$, $c_j$, $1\le j \le r$, $c_{r+1}=0$, set
\bea \nn
g_j(n,t) &=& \sum_{\ell=0}^j c_{j-\ell} \spr{\delta_n}{H(t)^\ell \delta_n},\\ \label{todaghsp}
h_j(n,t) &=& 2 a(n,t) \sum_{\ell=0}^j c_{j-\ell}  \spr{\delta_{n+1}}{H(t)^\ell
\delta_n} + c_{j+1},
\eea
and consider the Lax operator
\begin{equation}  \label{btgptdef}
P_{2r+2}(t) = -H(t)^{r+1} + \sum_{j=0}^r ( 2a(t) g_j(t) S^+ -h_j(t)) H(t)^{r-j} +
g_{r+1}(t),
\end{equation}
where $S^\pm f(n) = f(n\pm1)$. Restricting to the two-dimensional nullspace $\Ker(\tau(t)
-z)$, $z\in\C$, of $\tau(t)-z$, we have the following representation of $P_{2r+2}(t)$
\begin{equation} \label{btqptFG}
P_{2r+2}(t)\Big|_{\Ker(\tau(t)-z)} =2a(t) G_r(z,t) S^+ - H_{r+1}(z,t),
\end{equation}
where $G_r(z,n,t)$ and $H_{r+1}(z,n,t)$ are monic polynomials in $z$ of the
type
\bea \nn
G_r(z,n,t) &=& \sum_{j=0}^r z^j g_{r-j}(n,t),\\ \label{fgdef}
H_{r+1}(z,n,t) &=& z^{r+1} + \sum_{j=0}^r z^j h_{r-j}(n,t) - g_{r+1}(n,t).
\eea
A straightforward computation shows that the Lax equation
\begin{equation} \label{laxp}
\frac{d}{dt} H(t) -[P_{2r+2}(t), H(t)]=0, \qquad t\in\R,
\end{equation}
is equivalent to
\bea \nn
\tl_r (a(t), b(t))_1 &=& \dot{a}(t) -a(t) \Big(g_{r+1}^+(t) -
g_{r+1}(t) \Big)=0,\\ \label{tlrabo}
\tl_r (a(t), b(t))_2 &=& \dot{b}(t) - \Big(h_{r+1}(t) -h_{r+1}^-(t) \Big)=0,
\eea
where the dot denotes a derivative with respect to $t$ and $f^\pm(n)=f(n\pm 1)$.
Varying $r\in \N_0$ yields the Toda hierarchy
$\tl_r(a,b) =(\tl_r (a,b)_1, \tl_r (a,b)_2) =0$. We will always consider $r$ as
a fixed, but arbitrary, value.

We recall that the Lax equation (\ref{laxp}) implies existence of
a unitary propagator $U_r(t,s)$ such that the family of operators
$H(t)$, $t\in\R$, are unitarily equivalent, $H(t) = U_r(t,s) H(s) U_r(s,t)$.
This also implies the basic existence and uniqueness theorem for the Toda hierarchy
(see, e.g., \cite{tist}, \cite{ttkm}, or \cite[Section~12.2]{tjac}).

\begin{theorem} \label{thmexistandunique}
Suppose $(a_0,b_0) \in M = \ell^\infty(\Z) \oplus \ell^\infty(\Z)$. Then there exists a unique
integral curve $t \mapsto (a(t),b(t))$ in $C^\infty(\R,M)$ of the Toda hierarchy, that is,
$\tl_r(a(t),b(t))=0$, such that $(a(0),b(0)) = (a_0,b_0)$.
\end{theorem}

\noindent
Finally, we recall the following result from \cite{emtist} (compare also  \cite{tist}),
which says that solutions which are asymptotically close to a background
solution at the initial time stay close for all time.

\begin{lemma} \label{lemtsr}
Suppose $a(n,t)$, $b(n,t)$ and $\bar a(n,t)$, $\bar b(n,t)$ are two arbitrary 
bounded solutions of the Toda hierarchy satisfying
(\ref{H t2}) for one $t_0 \in \R$, then (\ref{H t2}) holds for all 
$t \in \R$, that is, 
\be \label{H t2}
\sum_{n \in \Z} w(n) \Big(|a(n,t) - \bar a(n,t)| + |b(n,t) - \bar b(n,t)| \Big) 
< \infty,
\ee
where $w(n)>0$.
\end{lemma}

\section{Quasi-periodic finite-gap solutions}
\label{secqp}

As a preparation for our next section we first need to recall some facts on
quasi-periodic finite-gap solutions (again see \cite{bght} or \cite{tjac}).

Let $\M$ be the Riemann surface associated with the following function
\begin{equation}
\Rg{z}= -\prod_{j=0}^{2g+1} \sqrt{z-E_j}, \qquad
E_0 < E_1 < \cdots < E_{2g+1},
\end{equation}
where $g\in \N$ and $\sqrt{.}$ is the standard root with branch cut along $(-\infty,0)$. 
$\M$ is a compact, hyperelliptic Riemann surface of genus $g$.
A point on $\M$ is denoted by 
$p = (z, \pm \Rg{z}) = (z, \pm)$, $z \in \C$, or $p = \infty_{\pm}$, and
the projection onto $\C \cup \{\infty\}$ by $\pi(p) = z$. The sets 
$
\Pi_{\pm} = \{ (z, \pm \Rg{z}) \mid z \in \C\backslash
\bigcup_{j=0}^g[E_{2j}, E_{2j+1}]\} \subset \M
$
are called upper, lower sheet, respectively.

Now pick $g$ numbers (the Dirichlet eigenvalues)
\be
(\hat{\mu}_j)_{j=1}^g = (\mu_j, \sigma_j)_{j=1}^g
\ee
whose projections lie in the spectral gaps, that is, $\mu_j\in[E_{2j-1},E_{2j}]$.
Associated with these numbers is the divisor $\dimuz$ which
is one at the points $\hat{\mu}_j$  and zero else. Using this divisor we
introduce
\begin{align} \nn
\ulz(p,n,t) &= \hAmap(p) - \hamap(\dimuz) - n\ul{\hat A}_{\infty_-}(\infty_+)
+ t\ul{U}_s - \hvrc \in \C^g, \\ \label{ulz}
\ulz(n,t) &= \ulz(\infty_+,n,t),
\end{align}
where $\vrc$ is the vector of Riemann constants, 
$\ul{U}_s$ the $b$-periods of the Abelian differential
$\Omega_s$ defined below, and $\Amap$ ($\amap$)
is Abel's map (for divisors). The hat indicates that we
regard it as a (single-valued) map from $\hat{\M}$ (the fundamental polygon
associated with $\M$) to $\C^g$.
We recall that the function $\theta(\ulz(p,n,t))$ has precisely $g$ zeros
$\hmu_j(n,t)$ (with $\hmu_j(0,0)=\hmu_j$), where $\theta(\ul{z})$ is the
Riemann theta function of $\M$.

Taking a stationary solution of $\tl_g$ with constants $c_j$, $1\le j \le g$, as
initial condition for another equation $\widehat{\tl}_s$ with constants
$\hat{c}_j$, $1\le j \le s$, in the Toda hierarchy (\ref{tlrabo}) one obtains
the quasi-periodic finite gap solutions of the Toda hierarchy given by
(see \cite[Sections~13.1, 13.2]{tjac})
\begin{align} \nn
a_q(n,t)^2 &= \ti{a}^2 \frac{\theta(\ulz(n+1,t)) \theta(\ulz(n-1,t))}{\theta(
\ulz(n,t))^2},\\ \label{imfab}
b_q(n,t) &= \tilde{b} + \sum_{j=1}^g c_j(g)
\frac{\partial}{\partial w_j} \ln\Big(\frac{\theta(\ul{w} +
\ulz(n,t)) }{\theta(\ul{w} + \ulz(n-1,t))}\Big) \Big|_{\ul{w}=0}.
\end{align}
The constants $\ti{a}$, $\tilde{b}$, $c_j(g)$ depend only on the Riemann surface
(see \cite[Section~9.2]{tjac}).

Introduce
\begin{align} \nn \label{BAfthetarep}
\phi_q(p,n,t) &= C(n,t) \frac{\theta (\ulz(p,n+1,t))}{\theta (\ulz(p,n,t))}
\exp \Big( \int_{p_0}^p \om_{\infty_+,\infty_-} \Big),\\  \quad
\psi_q(p,n,t) &= C(n,0,t) \frac{\theta (\ulz(p,n,t))}{\theta(\ulz (p,0,0))}
\exp \Big( n \int_{p_0}^p \om_{\infty_+,\infty_-} + t\int_{p_0}^p \Omega_s
\Big),
\end{align}
where $C(n,t)$, $C(n,0,t)$ are real-valued,
\begin{equation}
C(n,t)^2 = \frac{\theta(\ulz(n-1,t))}{\theta(\ulz(n+1,t))}, \qquad
C(n,0,t)^2 = \frac{ \theta(\ulz(0,0)) \theta(\ulz(-1,0))}
{\theta (\ulz (n,t))\theta (\ulz (n-1,t))},
\end{equation}
and the sign of $C(n,t)$ is opposite to that of $a_q(n,t)$.
$\om_{\infty_+,\infty_-}$ is the Abelian differential of the third kind with poles
at $\infty_+$ respectively $\infty_-$ and $\Omega_s$ is an Abelian differential
of the second kind with poles at $\infty_+$ respectively $\infty_-$ whose
Laurent expansion is given by the coefficients $(j+1)\hat{c}_{s-j}$ associated with
$\widehat{\tl}_s$ (see \cite[Sections~13.1, 13.2]{tjac}). Then
\begin{align}  \nonumber
\tau_q(t) \psi_q(p,n,t) &= \pi(p) \psi_q(p,n,t), \\ \nonumber
\frac{d}{dt} \psi_q(p,n,t) &= 2 a_q(n,t) \hat G_s(p,n,t) 
\psi_q(p,n+1,t) - \hat H_{s+1}(p,n,t) \psi_q(p,n,t) \\ \label{d/dt psi 2}
&= \hat P_{q,2s+2}(t) \psi_q(p,n,t),
\end{align}
where we use the hat to distinguish the quantities associated with $\widehat{\tl}_s$
from those associated with $\tl_g$.

The two branches $\psi_{q,\pm}(z,n,t)= \psi_q(p,n,t)$, $p=(z,\pm)$, of the Baker-Akhiezer
function are linearly independent away from the branch points and their Wronskian
is given by
\begin{equation}  \nonumber
W_q(\psi_{q,-}(z), \psi_{q,+}(z)) = \frac{R^{1/2}_{2g+2}(z)}{\prod_{j=1}^g (z-\mu_j)}.
\end{equation}
Here $W_q(f,g)=a_q(n)(f(n)g(n+1)-f(n+1)g(n))$ is the usual modified Wronskian.

It is well known that the spectrum of $H_q(t)$ is time independent and
consists of $g+1$ bands
\begin{equation}
\sig(H_q) = \bigcup_{j=0}^g [E_{2j},E_{2j+1}].
\end{equation}
For further information and proofs we refer to \cite[Chapter~9]{tjac}.

Finally, let us renormalize the Baker-Akhiezer function
\be \label{norm tilde}
\tilde \psi_q(p,n,t) = \frac{\psi_q(p,n,t)}{\psi_q(p,0,t)}
\ee
such that $\tilde \psi_q(p,0,t)=1$ and let us define $\alpha_s(p,t)$ via
\be \label{defalpha}
\exp\big(\alpha_s(p,t)\big)= \psi_q(p,0,t) =
C(0,0,t) \frac{\theta (\ulz(p,0,t))}{\theta(\ulz (p,0,0))}
\exp \Big( t\int_{p_0}^p \Omega_s \Big).
\ee

\section{Commutation methods and $N$-soliton solutions}
\label{secdc}

In this section we investigate commutation methods when applied
to a quasi-periodic finite-gap background solution. In particular, we
compute the phase shift (in the Jacobian variety) introduced by the
solitons. This can be found for the
case of one-dimensional Schr\"odinger operators in \cite{gesv} (see also \cite{gss}
for the elliptic case). The case of Jacobi operators seems to be missing and hence
we provide the corresponding results to fill this gap. We want to be rather brief
and refer to \cite{gtjc} or \cite[Ch.~11]{tjac} for further details in this connection.
Since the time $t$ does not play a role in this section, we will just omit it.

We start by inserting an eigenvalue using the single commutation method.  
Let $H_q$ be a quasi-periodic finite-gap operator and let $\psi_{q,\pm}(z,n)$ be
the branches of the Baker-Akhiezer function which are square summable near $\pm\infty$.
Fix $\lam_1 < \inf\sig(H_q)$, $\sigma \in [-1,1]$, define
\begin{equation}
u_{q,\sigma}(\lam_1,n) = \frac{1+\sigma}{2} \psi_{q,+}(\lam_1,n)
 + \frac{1-\sigma}{2} \psi_{q,-}(\lam_1,n),
\end{equation}
and let $H_{q,\sigma}$ be the (self-adjoint) commuted operator associated with  
\begin{equation} \label{sc}
(\tau_{q,\sigma}f)(n) = a_{q,\sigma}(n) f(n+1) + a_{q,\sigma}(n-1) f(n-1) +
b_{q,\sigma}(n) f(n),
\end{equation}
where (see \cite[Sect.~11.2]{tjac})
\begin{align} \nn
a_{q,\sigma}(n) &= -\frac{\sqrt{a_q(n) a_q(n+1) u_{q,\sigma}(\lam_1,n)
u_{q,\sigma}(\lam_1,n+2)}}{u_{q,\sigma}(\lam_1,n+1)}, \\ \nn
b_{q,\sigma}(n) &= \lam_1 -a_q(n) \bigg( \frac{u_{q,\sigma}(\lam_1,n)}
{u_{q,\sigma}(\lam_1,n+1)} + \frac{u_{q,\sigma}(\lam_1,n+1)}
{u_{q,\sigma}(\lam_1,n)} \bigg)\\ 
&= b_q(n) +\partial^* \frac{a_q(n)
u_{q,\sigma}(\lam_1,n)}{u_{q,\sigma}(\lam_1,n+1)}.
\end{align}
$H_q-\lam_1$ and $H_{q,\sigma}-\lam_1$ restricted to the 
orthogonal complements of their corresponding one-dimensional null-spaces are 
unitarily equivalent and hence
\begin{align*} \label{specsc}
\sig_{ac}(H_{q,\sigma}) &= \sig_{ac}(H_q), \qquad \sig_{sc}(H_{q,\sigma})
= \sig_{sc}(H_q) = \emptyset, \\
\sig_{pp}(H_{q,\sig}) &= \left\{ \begin{array}{cl} 
\{ \lam_1\}, & \sig \in (-1,1)\\ \emptyset, 
& \sig \in \{-1,1\} \end{array} \right.. 
\end{align*}

\begin{lemma} \label{lempsdc}
Let $H_q$ be a quasi-periodic finite-gap operator associated with the
Dirichlet divisor $\dimuz$ and let $H_{q,\sig}$, $-1<\sig<1$, be the commuted
operator associated with (\ref{sc}). Then we have
\begin{equation}
a_{q,\sig}(n) \sim a_{q,\pm1}(n) , \quad b_{q,\sig}(n) \sim b_{q,\pm1}(n) \qquad
\mbox{as } n\to \pm\infty,
\end{equation}
where $H_{q,\pm1}$ are the quasi-periodic finite-gap operators associated with the
Dirichlet divisors ${\dimuz}_{\pm1}$ defined via
\begin{equation}
\amap({\dimuz}_{\pm1}) = \amap(\dimuz) - \Amap(p_1) - \Amap(\infty_+), \qquad
p_1=(\lam_1,\pm).
\end{equation}
\end{lemma}

\begin{proof}
That $H_{q,\pm1}$ is associated with the divisor ${\dimuz}_{\pm1}$ is shown in
\cite[Sect.~11.4]{tjac} and the asymptotics follow since
$u_{q,\sigma}(\lam_1,n) \sim \frac{1\pm\sigma}{2} u_{q,\pm 1}(\lam_1,n)$ as
$n\to\mp\infty$.
\end{proof}

\noindent
Similarly, one obtains the following result for the double commutation method.
Let $\lam_1 \in \R \backslash \sigma_{ess}(H_q)$, 
define (see \cite[Sect.~11.6, (2.30)]{tjac})
\begin{align} \nn
c_{q,\gam}(\lam_1,n) &= \frac{1}{\gam} + \sum_{j=-\infty}^n \psi_{q,-}(\lam_1,j)^2\\  \label{c_gam}
& = \frac{1}{\gam} + W_{q,n}(\psi_{q,-}(\lam_1), \dot \psi_{q,-}(\lam_1))
= \frac{1}{\gam} + \psi_{q,-}(\lam_1,n)^2 \dot \phi_{q,-}(\lam_1,n),
\quad \gam\ne 0,
\end{align}
and let $H_{q,\gam}$ be the doubly commuted operator associated with  
\begin{align}  \nn
a_{q,\gam}(n) &= a_q(n) \frac{\sqrt{c_{q,\gam}
(\lam_1,n-1)c_{q,\gam}(\lam_1,n+1)}}{c_{q,\gam}(\lam_1,n)},\\ \label{dc}
b_{q,\gam}(n) &= b_q(n) - \partial^* 
\frac{a_q(n) \psi_{q,-}(\lam_1,n) \psi_{q,-}(\lam_1,n+1)}{c_{q,\gam}(\lam_1,n)}.
\end{align}

Then

\begin{lemma}
Let $H_q$ be a quasi-periodic finite-gap operator associated with the
Dirichlet divisor $\dimuz$ and let $H_{q,\gam}$, $0<\gam<\infty$, be the doubly commuted
operator associated with (\ref{dc}). Then we have
\begin{align} \nonumber
a_{q,\gam}(n) &=
\left\{ \begin{array}{c@{\qquad\mbox{as }}l}
a_q(n)(1+O(w(\lam_1)^{2n}) & n \to - \infty\\
a_{q,\infty}(n)(1+O(w(\lam_1)^{-2n}) & n \to + \infty
\end{array}\right., \\
b_{q,\gam}(n) &=
\left\{ \begin{array}{c@{\qquad\mbox{as }}l}
b_q(n)(1+O(w(\lam_1)^{2n}) & n \to - \infty\\
b_{q,\infty}(n)(1+O(w(\lam_1)^{-2n}) & n \to + \infty
\end{array}\right.,
\end{align}
where $w(z)= \exp(\int_{p_0}^{(z,+)} \om_{\infty_+,\infty_-})$ is the quasi-momentum map
and $H_{q,\infty}$ is the quasi-periodic finite-gap operator associated
with the Dirichlet divisor ${\dimuz}_\infty$ defined via
\begin{equation}
\amap({\dimuz}_{\lam_1}) = \amap(\dimuz) + 2 \Amap(\hat\lam_1), \qquad
\hat\lam_1=(\lam_1, +).
\end{equation}
\end{lemma}

\begin{proof}
Since the double commutation method can be obtained via two single
commutation steps (see \cite[Sect.~11.5]{tjac}), the result is a consequence of
our previous lemma. The asymptotics follow from (\ref{dc}) and (\ref{BAfthetarep}).
\end{proof}

\noindent
Clearly, if we add $k$ eigenvalues $\lam_1,\dots,\lam_k$, then the asymptotics at
$+\infty$ are given by the quasi periodic operator associated with the Dirichlet divisor
\begin{equation}
\amap({\dimuz}_{\lam_1,\dots,\lam_k}) = \amap(\dimuz) +
2 \sum_{j=1}^k \Amap(\hat\lam_j), \qquad \hat\lam_j=(\lam_j, +).
\end{equation}
In particular, by choosing at least one eigenvalue in each gap, we can attain
any prescribed asymptotics in the given isospectral class by
Lemma~9.1 in \cite{tjac}.

\begin{remark}
If $a_q(n,t)$, $b_q(n,t)$ is a quasi-periodic solution of the Toda hierarchy and
$\psi_q(p,n,t)$ is the corresponding time dependent Baker-Akhiezer function,
then $a_{q,\gam}(n,t)$, $b_{q,\gam}(n,t)$ is a solution of the Toda hierarchy 
which is centered at
\be
2 \alpha(\lam_1)(n - v(\lam_1) t) + \ln(\gam) = 0,
\ee
where
\be
\alpha(\lam_1)= \re \int_{p_0}^{(\lam_1,-)}\!\! \om_{\infty_+,\infty_-}, \qquad
v(\lam_1)=  -\frac{1}{\alpha(\lam_1)} \re \int_{p_0}^{(\lam_1,-)}\!\! \Omega_s.
\ee
\end{remark}

\section{Scattering theory}
\label{secst}

In this section we review scattering theory for Jacobi operators with steplike
quasi-periodic finite-gap background in the same isospectral class following
\cite{emtqst}. Our only new result in this section will be a full description of
the effect of the double commutation method on the scattering data in
Lemma~\ref{lemscatdc}.

More precisely, we will take two quasi-periodic
finite-gap operators $H_q^\pm$ associated with the sequences $a_q^\pm$, $b_q^\pm$
in the same isospectral class,
\be
\sig(H_q^+) = \sig(H_q^-) \equiv \Sigma = \bigcup_{j=0}^g [E_{2j},E_{2j+1}],
\ee
but with possibly different Dirichlet data $\{\hat{\mu}_j^\pm\}_{j=1}^g$. We will
add $\pm$ as a superscript to all data introduced in Section~\ref{secqp} to
distinguish between the corresponding data of $H_q^+$ and $H_q^-$. 
To avoid excessive sub/superscripts we abbreviate
\be
\psi_q^\pm(z, n)= \psi_{q,\pm}^\pm(z, n) \quad\mbox{and}\quad
\bpsi_q^\pm(z, n)= \psi_{q,\mp}^\pm(z, n),
\ee
that is, $\psi_q^\pm(z, n)$ is the solution of $H_q^\pm$ decaying near $\pm\infty$ and
$\bpsi_q^\pm(z, n)$ is the solution of $H_q^\pm$ decaying near $\mp\infty$.
Note that for $\lam\in\Sigma$ we have $\bpsi_q^\pm(\lam,n)= \overline{\psi_q^\pm(\lam,n)}$.

Let $a(n)$, $b(n)$ be sequences satisfying
\be \label{hypo}
\sum_{n = 0}^{\pm \infty} |n| \Big(|a(n) - a_q^\pm(n)| + |b(n) - b_q^\pm(n)| \Big)
< \infty
\ee
and denote the corresponding operator by $H$.

\begin{theorem} \label{thmjost}
Assume (\ref{hypo}). Then there exist solutions
$\psi_{\pm}(z, .)$, $z \in \C$, of $\tau \psi = z \psi$ satisfying
\begin{equation}                         \label{perturbed sol}
\lim_{n \rightarrow \pm \infty}
|w(z)^{\mp n} (\psi_{\pm}(z, n) - \psi_q^\pm(z, n))| = 0,
\end{equation}
where $w(z)= \exp(\int_{p_0}^{(z,+)} \om_{\infty_+,\infty_-})$ is the quasi-momentum map.
\end{theorem}

\begin{theorem} 
Assume (\ref{hypo}). Then we have $\sig_{ess}(H)=\Sigma$, the
point spectrum of $H$ is finite and confined to the spectral gaps of
$H_q^\pm$, that is, $\sig_p(H) = \{ \rho_j\}_{j=1}^q \subset \R\backslash\Sigma$.
Furthermore, the essential spectrum of $H$ is purely absolutely continuous.
\end{theorem}

\noindent
Using the fact that $\psi_q^\pm(p,n)$ form an orthonormal basis for
$L^2(\partial \Pi_+,d\omega^\pm)$, where
\be \label{domega}
d\omega^\pm = \frac{\prod_{j=1}^g(\pi-\mu_j^\pm)}{R^{1/2}_{2g+2}} d\pi,
\ee
we can define  
\begin{equation}
K_{\pm}(n,m) = 2 \re \int_\Sigma \psi_\pm(\lam,n) \psi_q^\pm(\lam, m)
d\omega^\pm 
\end{equation}
implying
\be \label{transop}
\psi_\pm(z,n) = \sum_{m=n}^{\pm \infty} K_\pm(n,m) \psi_q^\pm(z, m).
\ee
Next we define the coefficients of the scattering matrix via the scattering relations
\be
\psi_\mp (\lam,n) = \alpha_\pm(\lam) \overline{\psi_{\pm}(\lam,n)} 
+ \beta_\pm(\lam) \psi_{\pm}(\lam,n), \qquad \lam \in \Sigma,
\ee
where
\begin{align} \label{alpha quasi}
\alpha_\pm(\lam) &= \frac{W(\psi_\pm(\lam), \psi_\mp(\lam))}
{W(\psi_\pm(\lam), \overline{\psi_\pm(\lam)})} 
= \frac{\prod_{j=1}^g (\lam-\mu_j^\pm)}{\Rg{\lam}} W(\psi_-(\lam), \psi_+(\lam)), \\ \nn
\beta_\pm(\lam) &= \frac{W(\psi_\mp(\lam), \overline{\psi_\pm(\lam)})}
{W(\psi_\pm(\lam), \overline{\psi_\pm(\lam)})} 
= \mp \frac{\prod_{j=1}^g (z-\mu_j^\pm)}{\Rg{z}}W(\psi_\mp(\lam), \overline{\psi_\pm(\lam)}), 
\end{align}
and $W_n(f,g)=a(n)(f(n)g(n+1)-f(n+1)g(n))$ denotes the Wronskian.
Transmission $T_\pm(\lam)$ and reflection $R_{\pm}(\lam)$ 
coefficients are then defined by
\be
T_\pm(\lam) = \alpha_\pm^{-1}(\lam), \qquad
R_{\pm}(\lam) = \frac{\beta_\pm(\lam)}{\alpha_\pm(\lam)} = 
\frac{W(\psi_\mp(\lam), \overline{\psi_\pm(\lam)})}{W(\psi_\pm(\lam), \psi_\mp(\lam))}.
\ee

\noindent
The norming constants $\gam_{\pm, j}$ corresponding to $\rho_j \in \sigma_p(H)$ 
are given by
\be \label{norming}
\gam_{\pm, j}^{-1}=\sum_{n \in \Z}|\psi_\pm(\rho_j, n)|^2, \qquad 1 \leq j \leq q.
\ee
Note that $\gam_{\pm, j}=0$ if $\rho_j$ coincides with a pole $\hat \mu_\ell^\pm \in \Pi_\pm$  
of $\psi_\pm(z, .)$. To avoid this, one could remove the poles by introducing
$\hat \psi_\pm(z, .)$ as we did in \cite{emtqst}. Since this normalization cancels out in the Gel'fand-Levitan-Marchenko equation and unnecessarily complicates the formulas below, we will
allow zero norming constants. 
 
Moreover, $\psi_\pm(\rho_j, .)= c_j^\pm \psi_\mp(\rho_j, .)$ with $c_j^+ c_j^-=1$.

\begin{lemma}
The coefficients $T_\pm(\lam)$, $R_\pm(\lam)$ are bounded for $\lam \in \Sigma$, continuous for
$\lam \in \Sigma$ except at possibly the band edges $E_j$, and fulfill
\begin{align}
T_+(\lam) \overline{T_-(\lam)} + |R_\pm(\lam)|^2 &= 1, \qquad \lam \in \Sigma,\\
T_\pm(\lam) \overline{R_\pm(\lam)} + \overline{T_\pm(\lam)} R_\mp(\lam) &= 0, \qquad \lam \in \Sigma.
\end{align}
In particular,
\be
|T_\pm(\lam)|^2  \prod_{j=1}^g\frac{\lam - \mu_j^\pm}{\lam - \mu_j^\mp} 
+ |R_\pm(\lam)|^2 =1,
\ee
and hence $|R_\pm(\lam)|^2\le 1$ with equality only possibly at the band edges $\{E_j\}$.
The transmission coefficients $T_\pm(\lam)$ have a meromorphic continuation to $\C \backslash \Sigma$ 
with simple poles at $\mu_j^\pm$ if $\hat\mu_j^\pm \in \Pi_\mp$ and simple poles at $\rho_j$,
\be \label{res T}
(\res_{\rho_j}T_\pm(\lam))^2 = \gam_{+,j}\gam_{-,j} \frac{R_{2g+2}(\rho_j)}{\prod_{l=1}^g (\rho_j-\mu_l^\pm)^2}.
\ee
In addition, $T_\pm(z) \in \R$ as $z \in \R \backslash \Sigma$ and
\be \label{T infty}
T_\pm(\infty) = \prod_{n=-\infty}^{-1} \frac{a(n)}{a_q^-(n)} \prod_{n=0}^\infty \frac{a(n)}{a_q^+(n)}.
\ee
\end{lemma}

The sets  
\be
S_{\pm}(H) = \{R_{\pm}(\lam), \lam \in \Sigma; \, (\rho_j, \gam_{\pm, j}), 
1 \leq j \leq q\}
\ee
are called left/right scattering data for $H$.

\begin{theorem}
The kernel $K_{\pm}(n,m)$ of the transformation operator satisfies
the Gel'fand-Levitan-Marchenko equation
\be   \label{glm1 q}
K_{\pm}(n,m) + \sum_{l=n}^{\pm \infty}K_{\pm}(n,l)F^{\pm}(l,m) = 
\frac{\delta(n,m)}{K_{\pm}(n,n)},  \qquad \pm m \geq \pm n,
\ee
where
\begin{align} \label{glm2 q}
F^\pm(m,n)&= 2\re\int_{\Sigma}R_\pm(\lam) \psi_q^\pm(\lam,m) 
\psi_q^\pm(\lam,n) d\omega^\pm   + \sum_{j=1}^q \gam_{\pm,j} 
\psi_q^\pm(\rho_j,n) \psi_q^\pm(\rho_j,m).
\end{align}
Note that the apparent poles $\mu_\ell^\pm$ cancel with the zeros of 
$d\omega^\pm$ and $\gam_{\pm,j}$ at these points.
\end{theorem}

\noindent
The operator $H$ can be uniquely reconstructed from $S_\pm(H)$ by solving the 
Gel'fand-Levitan-Marchenko equation. We refer to \cite{emtqst} for further details.

Finally, we come to our principal new result in this section and 
investigate the connection with the double commutation method.
The scattering data of the operators $H$, $H_\gam$ are related as follows.

\begin{lemma} \label{lemscatdc}
Let $H$ be a given Jacobi operator satisfying (\ref{hypo}) and choose
$\rho_{q+1}\in\R\backslash\sig(H)$, $\gam>0$. Then the doubly commuted 
operator $H_\gam$, $\gam > 0$, defined via $\psi_-(\rho_{q+1})$ as in
Section~\ref{secdc}, satisfies
\begin{equation}
a_\gam(n) \sim \left\{ \begin{array}{c@{\quad\mbox{as }}l}
a_q^-(n) & n \to -\infty\\
a_q^\infty(n) & n \to +\infty
\end{array}\right., \qquad
b_\gam(n) \sim \left\{ \begin{array}{c@{\quad\mbox{as }}l}
b_q^-(n) & n \to -\infty\\
b_q^\infty(n) & n \to +\infty
\end{array}\right.,
\end{equation}
such that (\ref{hypo}) still holds,
where $H_q^\infty$ is associated with Dirichlet data given by
\begin{equation}
\amap(\di_{\ul{\mu}^\infty}) = \amap(\di_{\ul{\mu}^+}) + 2\Amap(\hat{\rho}_{q+1}).
\end{equation}
It has the scattering data 
\begin{align}
R_{-,\gam}(\lam) &= R_-(\lam), \\
R_{+,\gam}(\lam) &= \frac{\theta(\ulz^\infty(\hat\lam,0))}{\theta(\ulz^+(\hat\lam,0))}
\frac{\theta(\ulz^+(\hat\lam^*,0))}{\theta(\ulz^\infty(\hat\lam^*,0))}
B(\lam,\rho_{q+1})^2 R_+(\lam), \\
T_{+,\gam}(z) &= C \, 
\frac{\theta(\ulz^+(\hat z^*,0))}{\theta(\ulz^\infty(\hat z^*,0))} 
B(z,\rho_{q+1}) T_+(z),\\
T_{-,\gam}(z) &= \frac{1}{C} \,
\frac{\theta(\ulz^\infty(\hat z,0))}{\theta(\ulz^+(\hat z,0))} B(z,\rho_{q+1}) 
T_-(z),
\end{align}
where 
\be
B(z,\rho)= \exp\bigg(-\int_{E(\rho)}^{(\rho,+)} \!\!\! \om_{\hat z \hat z^*}\bigg) 
\ee
is the Blaschke factor and $\hat\lam=(\lam,+)$, $\hat z=(z,+)$. 
The constant $C$ is given by
\be
C = \sqrt{\frac{\theta(\ulz^\infty(\infty_+,0))\theta(\ulz^\infty(\infty_-,0))}
{\theta(\ulz^+(\infty_+,0)) \theta(\ulz^+(\infty_-,0))}} > 0.
\ee
The norming constants $\gam_{-,j}$ corresponding to $\rho_j \in \sig_p(H)$, $j=1,\dots,q$,
(cf.\ (\ref{norming})) remain unchanged except for an additional eigenvalue
$\rho_{q+1}$ with norming constant $\gam_{-,q+1}=\gam$.
The norming constants $\gam_{+,j,\gam}$, $j=1,\dots,q$, are given by
\be
\gam_{+,j,\gam} = \frac{1}{C^2}
\left(\frac{\theta(\ulz^\infty(\hat\rho_j,0))}{\theta(\ulz^+(\hat\rho_j,0))}\right)^2
B(\rho_j,\rho_{q+1})^2 \gam_{+,j},  \qquad \hat\rho_j=(\rho_j,+), 
\ee
and the additional norming constant $\gam_{+,q+1,\gam}$ reads
\be
\gam_{+,q+1,\gam} = C^2 \frac{\prod_{j=1}^g(\rho_{q+1}-\mu_j^+)^2}{\gam R_{2g+2}(\rho_{q+1})}
\bigg( \frac{\theta(\ulz^+(\hat \rho_{q+1}^*,0))}{\theta(\ulz^\infty(\hat \rho_{q+1}^*,0))}
T_+(\rho_{q+1}) \res_{z=\rho_{q+1}} B(z,\rho_{q+1}) \bigg)^2.
\ee
\end{lemma}

\begin{proof}
First we show that (\ref{hypo}) still holds. Note that
\begin{align*}
a_\gam(n) &=
\left\{ \begin{array}{c@{\qquad\mbox{as }}l}
a(n)(1+O(w(\rho_{q+1})^{2n}) & n \to - \infty\\
a_\infty(n)(1+O(w(\rho_{q+1})^{-2n}) & n \to + \infty
\end{array}\right., \\
b_\gam(n) &=
\left\{ \begin{array}{c@{\qquad\mbox{as }}l}
b(n)(1+O(w(\rho_{q+1})^{2n}) & n \to - \infty\\
b_\infty(n)(1+O(w(\rho_{q+1})^{-2n}) & n \to + \infty
\end{array}\right..
\end{align*}
Hence the asymptotics near $-\infty$ are clearly unchanged and for $+\infty$ it suffices to
check $\gam=\infty$. By Lemma~\ref{lemgf} below,
\begin{align*}
\frac{c_\infty(\rho_{q+1}, n+1)}{c_\infty(\rho_{q+1}, n)} = 
\frac{c_\infty(\rho_{q+1}, n+1)\psi_+(\rho_{q+1},n+1)}{c_\infty(\rho_{q+1}, n)\psi_+(\rho_{q+1},n+1)} =
\frac{c_{q,\infty}^+(\rho_{q+1},n+1)}{c_{q,\infty}^+(\rho_{q+1},n)}(1+ C(n)),
\end{align*}
where $\sum_{n\in \N} n |C(n)|<\infty$. Therefore
\be
a_\infty(n) = a(n) \frac{\sqrt{c_{\infty}
(\rho_{q+1},n-1)c_{\infty}(\rho_{q+1},n+1)}}{c_{\infty}(\rho_{q+1},n)} \to
a_{q,\infty}(n)
\ee
such that $\sum_{n\in \N} n |a_\infty(n) - a_{q,\infty}(n)|< \infty$ and similarly for $b_\infty(n)$.

Now we turn to the scattering data. By \cite[Lemma 11.19]{tjac}, the Jost solutions 
$\psi_{\pm,\gam}(z,n)$ of $H_\gam$ are up to a constant given by 
\begin{equation}
u_{\pm, \gam}(z,n) = 
\frac{c_{\gam}(\rho_{q+1},n) \psi_\pm(z,n) - \frac{1}{z-\rho_{q+1}} \psi_-(\rho_{q+1},n)
W_{n-1}(\psi_-(\rho_{q+1}),\psi_\pm(z))}
{\sqrt{c_{\gam}(\rho_{q+1},n-1)c_{\gam}(\rho_{q+1},n)}}.
\end{equation}
Since this constant is equal to $1$ for $\psi_{-,\gam}(z,n)$ the fact that
$R_-$ is unchanged follows from its definition and \cite[(11.107)]{tjac}. 
The transmission coefficients are reconstructed from $R_-(\lam)$ using
\cite[Theorem~3.6]{emtqst} and $R_{+,\gam}(\lam)$ follows then from 
\[
R_{+,\gam}(\lam) = - \frac{T_{-,\gam}(\lam)}{\overline{T_{-,\gam}(\lam)}}R_{-,\gam}(\lam).
\]
That the norming constants $\gam_{-,j}$ are unchanged follows from \cite[Lemma 11.14]{tjac}.
For $\gam_{+,j,\gam}$, $j=1,\dots,q+1$, we use (\ref{res T})
\[
\gam_{+,j,\gam} = \frac{\prod_{l=1}^g (\rho_j-\mu_l^\pm)^2 (\res_{\rho_j}T_{\pm,\gam}(\lam))^2}
{R_{2g+2}(\rho_j)\gam_{-,j}}.
\]
\end{proof}

\begin{remark}
If we choose $\rho=\rho_j\in\sig_p(H)$ and $\gam=-\gam_{-,j}$, then the eigenvalue
$\rho_j$ is removed from the spectrum and it is straightforward to see that an 
analogous result holds.
\end{remark}

\noindent
The following result used in the previous proof is of independent interest.

\begin{lemma} \label{lemgf}
Let $H$ be a given Jacobi operator satisfying (\ref{hypo}). Then for every $z\in\C\backslash\Sigma$
and every $k\in\N$ we have
\begin{equation} \label{eqdiffgf}
\psi_\mp(z,n)\psi_\pm(z,n+k) - \alpha_\pm(z) \psi_q^\pm(z,n+k) \bpsi_q^\pm(z,n) = (k+1)|w(z)|^k C_\pm(z,n),
\end{equation}
where $\sum_{\pm n \in\N} n |C_\pm(z,n)| < \infty$.

Similarly, one has
\begin{equation} \label{eqdiffgf 2}
c_\infty(z,n) \psi_\pm(z,n+k) - \alpha_\pm(z) c_{q,\infty}^\pm(z,n) \psi_q^\pm(z,n+k) = 
\tilde C_\pm(z,n,k),
\end{equation}
where $\sum_{\pm n \in\N} n |\tilde C_\pm(z,n,k)| < \infty$ and $c_\infty(z,n)$, $c_{q,\infty}^\pm(z,n)$ 
are defined as in (\ref{c_gam}).
\end{lemma}

\begin{proof}
The proof is an extension of \cite[Lemma 3.4]{mtqptr} and 
we will only consider the '$+$' case. First recall that the Green's functions
of $H_q^+$ and $H$ are given by
\[
G_q^+(z,n,m) = \frac{\bpsi_q^+(z,m) \psi_q^+(z,n)}{W_q(\bpsi_q^+(z),\psi_q^+(z))},
\quad G(z,n,m) = \frac{\psi_-(z,m) \psi_+(z,n)}{W(\psi_-(z),\psi_+(z))}, \quad n\ge m,
\]
respectively. Considering matrix elements in the second resolvent identity
$(H-z)^{-1}-(H_q^+-z)^{-1}= (H-z)^{-1}(H_q^+-H)(H_q^+-z)^{-1}$ we obtain
\[
C_+(z,n) = \frac{|w(z)|^{-k}}{k+1}
\sum_{m\in\Z} W(\psi_-(z),\psi_+(z)) G(z,n,m) (H_q^+-H) G_q^+(z,m,n+k)
\]
for $z\in\C\backslash\sig(H)$. Since the poles of $W(\psi_-(z),\psi_+(z)) G(z,n,m)$ at
$z\in\sig_p(H)$ are removable, the formula holds for all $z\in\C\backslash\Sigma$.
Estimating the right hand side using $|G(z,n,m)| \le const\, |w(z)|^{|n-m|}$ and
$|G_q^+(z,n+k,m)| \le const\, |w(z)|^{|n+k-m|}$ we obtain
\[
|C_+(z,n) | \le  \frac{C |w(z)|^{-k}}{k+1} \sum_{m\in\Z}
|w(z)|^{|n-m|+|n-m+k|} (2|a(m)-a_q^+(m)|+|b(m)-b_q^+(m)|).
\]
We split the sum into three parts
\begin{align*}
|w(z)|^{|n-m|+|n-m+k|} &=
\left\{ \begin{array}{cl}
|w(z)|^k |w(z)|^{2|n-m|} & \quad m < n,\\
|w(z)|^k  &  \quad n \leq m\leq n+k, \\
|w(z)|^k |w(z)|^{2|n-m+k|} & \quad m > n+k.
\end{array}\right. 
\end{align*}
Since (\ref{hypo}) holds for $c(m)=2|a(m)-a_q^+(m)|+|b(m)-b_q^+(m)|$ 
and we have $|w(z)|<1$ for $z\in\C\backslash\Sigma$, we can apply Lemma~\ref{le:app} 
to verify $\sum_{n \in\N} n |C_+(z,n)| < \infty$.
 
The claim (\ref{eqdiffgf 2}) is a consequence of (\ref{c_gam}) and (\ref{eqdiffgf})
\begin{align*}
c_\infty(z,n) \psi_+(z,n+k)^2 &= \sum_{j=-\infty}^n \psi_-(z,j)^2\psi_+(z,n+k)^2 \\
&= \alpha_+(z)^2 c_{q,\infty}^+(z,n) \psi_q^+(z,n+k)^2 + \tilde C_\pm(z,n,k)\psi_q^+(z,n+k),
\end{align*}
where $\tilde C_\pm(z,n,k)$ again can be estimated using Lemma~\ref{le:app}.
\end{proof}

\section{Inverse scattering transform}
\label{secist}

Let $a(n,t)$, $b(n,t)$ be a solution of the Toda hierarchy satisfying
\be \label{hypot}
\sum_{n = 0}^{\pm \infty} |n| \Big(|a(n,t) - a_q^{\pm}(n,t)| + |b(n,t) - b_q^{\pm}(n,t)| \Big)
< \infty.
\ee
Note that by Lemma~\ref{lemtsr} it suffices to check (\ref{hypot}) for one $t_0 \in \R$
(as background take $H_q^-(t)$ and insert $g$ eigenvalues such that the asymptotics
on the other side are given by $H_q^+(t)$).

Jost solutions, transmission and reflection coefficients depend now on an 
additional parameter $t \in \R$. 
The Jost solutions $\psi_{\pm}(z,n,t)$ are normalized such that 
\[
\tilde\psi_\pm(z,n,t)= \tilde\psi_{q}^\pm(z,n,t)\,(1 + o(1)) \quad
\mbox{as } n\to\pm\infty,
\]
where (c.f. (\ref{norm tilde})) 
\be
\tilde \psi_q^\pm (z,n,t) = \frac{\psi_q^\pm (z,n,t)}{\psi_q^\pm (z,0,t)}
=: \exp(- \alpha_s^\pm(z,t))\psi_q^\pm (z,n,t).
\ee
Moreover, we set
\be
\exp(\bar \alpha_s^\pm(z,t))=\bpsi_q^\pm(z,0,t).
\ee
Note that we have $\overline{\exp(\alpha_s^\pm(z,t))}=\exp(\bar \alpha_s^\pm(z,t))$ 
for $\lam\in\Sigma$.

Transmission and reflection coefficients are then defined via the normalized Jost solutions 
$\ti \psi_{\pm}(z,n,t)$. Moreover, 
\be
\sigma(H(t)) \equiv \sigma(H),  \qquad t \in \R.
\ee

To avoid the poles of the Baker-Akhiezer function, we will assume that none of the
eigenvalues $\rho_j$ coincides with a Dirichlet eigenvalue $\mu_k^\pm(0,0)$. This
can be done without loss of generality by shifting the initial time $t_0=0$ if
necessary. 

\begin{remark}
Due to this assumption there is no need to remove these poles for the definition
of $\gam_{\pm, j}$, as we did in \cite{emtqps}, \cite{emtqst}. Since the
Dirichlet eigenvalues rotate in their gap, the factor needed to remove the poles
would only unnecessarily complicate the time evolution of the norming constants.
Moreover, these factors would eventually cancel in the Gel'fand-Levitan-Marchenko
equation, which is the only interesting object from the inverse spectral point of
view in the first place. 
\end{remark}

\begin{lemma} \label{th: sol H t}
Let $(a(t),b(t))$ be a solution of the Toda hierarchy such that (\ref{hypo}) holds.
The functions 
\be
\psi_{\pm}(z,n,t)=\exp(\alpha_{s}^\pm(z,t)) \tilde\psi_{\pm}(z,n,t) 
\ee
satisfy
\be \label{H t 1}
H(t) \psi_\pm(z,n,t) = z \psi_\pm(z,n,t), \qquad
\frac{d}{dt} \psi_\pm(z,n,t) = \hat P_{2s+2}(t) \psi_\pm(z,n,t).
\ee
\end{lemma}

\begin{proof}
We proceed as in \cite{emtist}, \cite[Theorem 3.2]{tist}. 
The Jost solutions $\psi_{\pm}(z,n,t)$ are continuously differentiable with respect to
$t$ by the same arguments as for $z$ (compare \cite[Theorem 4.2]{emtqps})
and the derivatives are equal to the derivatives of the Baker-Akhiezer functions
as $n \rightarrow \pm \infty$.

For $z \in \rho(H)$, the solution $u_{\pm}(z,n,t)$ of (\ref{H t 1}) with initial condition 
$\psi_{\pm}(z,n,0) \in \ell_{\pm}^2(\Z)$ remains square summable near $\pm \infty$
for all $t \in \R$ (see \cite{ttkm} or \cite[Lemma 12.16]{tjac}), that is,
$u_{\pm}(z,n,t) = C_\pm(t) \psi_{\pm}(z,n,t)$. Letting $n\to\pm\infty$ we
see $C_\pm(t)=1$. The general result for all $z \in \C$ now follows from continuity.
\end{proof}

\noindent
This implies

\begin{theorem} \label{thm scat t}
Let $(a(t),b(t))$ be a solution of the Toda hierarchy such that (\ref{hypo}) holds.
The time evolution for the scattering data is given by
\begin{align} 
T_{\pm}(z, t) &=  T_{\pm}(z, 0) \exp(\alpha_{s}^\mp(z,t)-\bar \alpha_{s}^\pm(z,t)), \\ 
R_{\pm}(\lam, t) &=  R_{\pm}(\lam, 0) 
\exp(\alpha_{s}^\pm(\lam,t)-\bar \alpha_{s}^\pm(\lam,t)), \quad \lam \in \Sigma,\\ 
\gam_{\pm, j}(t) &=  \gam_{\pm, j}(0) \exp(2 \alpha_{s}^\pm(\rho_j,t)), 
\qquad 1 \leq j \leq q.
\end{align}
\end{theorem}

\begin{proof}
Since the Wronskian of two solutions satisfying (\ref{H t 1}) does not depend on 
$n$ or $t$ (see \cite{ttkm}, \cite[Lemma 12.15]{tjac}), we have 
\begin{align*}
T_{\pm}(z, t) &= \frac{W(\overline{\tilde\psi_{\pm}(z,t)}, \tilde\psi_{\pm}(z,t))}
{W(\tilde\psi_{\mp}(z,t), \tilde\psi_{\pm}(z,t))}  = 
\frac{\exp(\alpha_{s}^\mp(z,t)+\alpha_{s}^\pm(z,t))}
{\exp(\bar \alpha_{s}^\pm(z,t) + \alpha_{s}^\pm(z,t))}
\frac{W(\overline{\psi_{\pm}(z,t)}, \psi_{\pm}(z,t))}
{W(\psi_{\mp}(z,t), \psi_{\pm}(z,t))} \\ 
&= \exp(\alpha_{s}^\mp(z,t)-\bar \alpha_{s}^\pm(z,t)) T_{\pm}(z, 0).
\end{align*}
The result for $R_{\pm}(\lam, t)$ follows similarly.
The time dependence of $\gam_{\pm, j}(t)$ follows from
$\|\psi_\pm(\rho_j,.,t)\|= \|\hat U_s(t,0)\psi_\pm(\rho_j,.,0)\|= \|\psi_\pm(\rho_j,.,0)\|$.
\end{proof}

\begin{corollary} The quantity $T_{\pm}(\lam,t)\overline{T_{\mp}(\lam,t)} = 1 - |R_\pm(\lam,t)|^2$,
$\lam\in\Sigma$,  does not depend on $t$.
\end{corollary}

\noindent
Another straightforward consequence is:

\begin{theorem}
The time dependence of the kernel of the Gel'fand-Levitan-Marchenko equation is
given by
\begin{align*}
F^\pm(m,n,t) &=
2\re\int_\Sigma
R_\pm(\lambda,0) \psi_q^\pm(\lam, m,t) \psi_q^\pm(\lam, n, t) 
d\omega^\pm(0)\\
&\quad + \sum_{j=1}^q \gam_{\pm,j}(0) \psi_q^\pm(\rho_j, m,t) \psi_q^\pm(\rho_j, n,t).
\end{align*}
\end{theorem}

\begin{proof}
Just employ Theorem~\ref{thm scat t} to rewrite 
\begin{align*}  
F^\pm(m,n,t)&=\int_{\partial\Pi_+}R_\pm(p,t) \ti \psi_q^\pm(p,m,t) 
\ti \psi_q^\pm(p,n,t) d\omega^\pm(t) \\
&\quad + \sum_{j=1}^q \gam_{\pm,j}(t) 
\ti\psi_q^\pm(\rho_j,n,t) \ti\psi_q^\pm(\rho_j,m,t),
\end{align*}
where we also use that  
$\exp(\alpha_{s}^\pm(\lam,t) + \bar \alpha_{s}^\pm(\lam,t))=
\prod_{j=1}^g \frac{\lam - \mu_j^\pm(0,t)}{\lam - \mu_j^\pm(0,0)}$.
\end{proof}

\noindent
Finally we note (\cite{ttkm}, \cite[Section 14.5]{tjac})

\begin{lemma}
Let $(a(t),b(t))$ be a solution of the Toda hierarchy such that (\ref{hypo}) holds.
Choose $\rho\in\R\backslash\sig(H)$ and $\gam>0$. Then
$(a_\gam(t),b_\gam(t))$ defined via $\psi_-(\rho,n,t)$ using the double commutation method
is again a solution of the Toda hierarchy such that (\ref{hypo}), with $H_q^+(t)$ accordingly
changed, holds.
\end{lemma}

{\bf Acknowledgments.}
I.E. gratefully acknowledges the extraordinary hospitality of the Faculty of
Mathematics of the University of Vienna during two stays in 2006,
where parts of this paper were written.

\begin{appendix}

\section{Some estimates}

In the proof of Lemma~\ref{lemgf} we need the following elementary result. Consider a sequence
$c\in\ell(\Z)$ and abbreviate
\be
\|c\|_\infty = \sup_{n\in\Z} |c(n)|, \quad
\|c\|_1 = \sum_{n=0}^\infty |c(n)|, \quad
\|c\|_{1,1} = \sum_{n=1}^\infty n |c(n)|.
\ee

\begin{lemma}\label{le:app}
Suppose $w$ is some complex number with $|w|<1$ and $c\in\ell(\Z)$ satisfies
$\|c\|_\infty, \|c\|_{1,1} <\infty$. 

Then
$$
\| \sum_{m=0}^\infty c(n+m) w^m \|_{1,1} \le \frac{1}{1-|w|} \|c\|_{1,1}
$$
and
$$
\| \sum_{m=0}^\infty c(n-m) w^m \|_{1,1} \le \frac{1}{1-|w|} \|c\|_{1,1} + \frac{|w|}{(1-|w|)^2} \|c\|_1
+ \frac{|w|^2}{(1-|w|)^3} \|c\|_\infty.
$$
\end{lemma}

\begin{proof}
The first estimate follows from
\begin{align*}
\| \sum_{m=0}^\infty c(n+m) w^m \|_{1,1} &=\sum_{n=1}^\infty n \Big| \sum_{m=0}^\infty c(n+m)w^m \Big|\\
&\leq \sum_{n=0}^\infty \sum_{m=0}^\infty (n+m)|c(n+m)||w|^m \\
&= \sum_{m=0}^\infty \|c\|_{1,1}  |w|^m = \frac{1}{1-|w|} \|c\|_{1,1}.
\end{align*}
Similarly, the second follows from
\begin{align*}
\| \sum_{m=0}^\infty c(n-m) w^m \|_{1,1} &=
\sum_{n=0}^\infty n \Big| \sum_{m=0}^\infty c(n-m) w^m \Big| \\
& \le \sum_{m=0}^\infty \left( \frac{m(m-1)}{2} \|c\|_\infty + m \|c\|_1 + \|c\|_{1,1}\right) |w|^m.
\end{align*}
 \end{proof}

\end{appendix}

\end{document}